\documentclass{aa}

\usepackage[varg]{txfonts}
\usepackage{graphicx}
\usepackage{color} 
\usepackage{amsmath}
\usepackage{multirow} 
\bibpunct{(}{)}{;}{a}{}{,} 

\begin{document}

\title{Imaging the water snowline in a protostellar envelope with H$^{\sf13}$CO$^+$\thanks{Based on observations carried out with the IRAM NOEMA interferometer. IRAM is supported by INSU/CNRS (France), MPG (Germany), and IGN (Spain).}}

\author{Merel L.R. van 't Hoff\inst{1}
\and Magnus V. Persson\inst{2}
\and Daniel Harsono\inst{1}
\and Vianney Taquet\inst{1,3}
\and Jes K. J{\o}rgensen\inst{4}
\and Ruud Visser\inst{5}
\and Edwin A. Bergin\inst{6}
\and Ewine F. van Dishoeck\inst{1,7}}

\institute{Leiden Observatory, Leiden University, P.O. Box 9513, 2300 RA Leiden, The Netherlands \\e-mail: \texttt{vthoff@strw.leidenuniv.nl}
\and Department of Space, Earth and Environment, Chalmers University of Technology, Onsala Space Observatory, 439 92 Onsala, Sweden
\and INAF, Osservatorio Astrofisico di Arcetri, Largo E. Fermi 5, 50125 Firenze, Italy
\and Centre for Star and Planet Formation, Niels Bohr Institute \& Natural History Museum of Denmark, University of Copenhagen, {\O}ster Voldgade 5–7, 1350 Copenhagen K., Denmark
\and European Southern Observatory, Karl-Schwarzschild-Str. 2, 85748 Garching, Germany
\and Department of Astronomy, University of Michigan, 1085 S. University Ave., Ann Arbor, MI 48109-1107, USA
\and Max-Planck-Institut f\"ur Extraterrestrische Physik, Giessenbachstrasse 1, 85748 Garching, Germany }

\date{}

\abstract {Snowlines are key ingredients for planet formation. Providing observational constraints on the locations of the major snowlines is therefore crucial for fully connecting planet compositions to their formation mechanism. Unfortunately, the most important snowline, that of water, is very difficult to observe directly in protoplanetary disks due to its close proximity to the central star.} 
{Based on chemical considerations, HCO$^+$ is predicted to be a good chemical tracer of the water snowline, because it is particularly abundant in dense clouds when water is frozen out. This work aims to map the optically thin isotopologue H$^{13}$CO$^+$ toward the envelope of the low-mass protostar NGC1333-IRAS2A, where the snowline is at larger distance from the star than in disks. Comparison with previous observations of H$_2^{18}$O will show whether H$^{13}$CO$^+$ is indeed a good tracer of the water snowline.} 
{NGC1333-IRAS2A was observed using the NOrthern Extended Millimeter Array (NOEMA) at $\sim$ $0\farcs9$ resolution, targeting the H$^{13}$CO$^+$ $J=3-2$ transition at 260.255 GHz. The integrated emission profile was analyzed using 1D radiative transfer modeling of a spherical envelope with a parametrized abundance profile for H$^{13}$CO$^+$. This profile was validated with a full chemical model.}
{The H$^{13}$CO$^+$ emission peaks $\sim$2$^{\prime\prime}$ northeast of the continuum peak, whereas H$_2^{18}$O shows compact emission on source. Quantitative modeling shows that a decrease in H$^{13}$CO$^+$ abundance by at least a factor of six is needed in the inner $\sim$360~AU to reproduce the observed emission profile. Chemical modeling predicts indeed a steep increase in HCO$^+$ just outside the water snowline; the 50\% decrease in gaseous H$_2$O at the snowline is not enough to allow HCO$^+$ to be abundant. This places the water snowline at 225 AU, further away from the star than expected based on the 1D envelope temperature structure for NGC1333-IRAS2A. In contrast, DCO$^+$ observations show that the CO snowline is at the expected location, making an outburst scenario unlikely.}
{The spatial anticorrelation of the H$^{13}$CO$^+$ and H$_2^{18}$O emission provide a proof of concept that H$^{13}$CO$^+$ can be used as a tracer of the water snowline.}

\keywords{ISM: individual objects: NGC1333-IRAS2A -- ISM: molecules -- astrochemistry -- stars: protostars -- submillimeter: planetary systems}

\maketitle


\section{Introduction}

Water is probably the molecule that appeals most to our imagination, as it is essential for life as we know it. In star-forming regions, water plays an important role as coolant of the warm gas, as dominant carrier of oxygen, and as major constituent of icy grain mantles deep inside a planet-forming disk \citep[see e.g.,][for reviews]{Melnick2009,Bergin2012,vanDishoeck2013}. The transition from water being frozen out onto dust grains to being predominantly present in the gas phase occurs at the water snowline: the midplane radius at which 50\% of the water is in the gas phase, and 50\% is in ice. Since the selective freeze-out of the major oxygen carrier alters the elemental C/O-ratio in both gas and ice, the bulk chemical composition of planets depends on their formation location with respect to the freeze-out radius of water \citep[e.g.,][]{Oberg2011,Madhusudhan2014,Ali-Dib2014,Ali-Dib2017,Walsh2015,Mordasini2016,Eistrup2016,Booth2017}. In addition, planetesimal formation is thought to be significantly enhanced in this region \citep[e.g.,][]{Stevenson1988,Schoonenberg2017}. The exact location and time evolution of the water snowline in protostellar systems is thus a crucial ingredient in planet formation.  

Unfortunately, the water snowline is very hard to observe directly in protoplanetary disks. Because of the large binding energy of H$_2$O, the transition from ice to gas happens a few AU from the young star where the midplane temperature exceeds $\sim$~100--200~K (depending on the vapor pressure). For the nearest star forming regions, this would already require angular resolutions of $\lesssim 0\farcs01$. Emission originating from cold water (E$_{\rm{up}}$~<~100~K) has been detected in the disk around TW Hya with the \textit{Herschel Space Observatory} \citep{Hogerheijde2011,Zhang2013}, but the large \textit{Herschel} beam (10--45$\arcsec$) could not spatially resolve the snowline. Furthermore, the only thermal water lines that can be observed from the ground (except for the H$_2$O line at 183~GHz) are lines from the less abundant isotopologue H$_2^{18}$O. As such, even ALMA will have great difficulty locating the water snowline in protoplanetary disks.

An alternative approach is to use chemical imaging, as has been done for the CO snowline. Due to CO being highly volatile, its snowline is located tens of AU from the central star (around $\sim$20~K; see e.g. \citealt{Burke2010}). Although this is far enough to be spatially resolved with ALMA, locating it directly remains difficult; since CO line emission is generally optically thick, it does not reveal the cold disk midplane. Reactions with gaseous CO are the main destruction route for N$_2$H$^+$, so N$_2$H$^+$ is expected to be abundant only when CO is frozen out. Taking simple chemical considerations into account, an upper limit for the CO snowline location can therefore be derived from N$_2$H$^+$ emission \citep{Aikawa2015,vantHoff2017}, as has been done for the disks around TW Hya and HD 163296 \citep{Qi2013,Qi2015}, and several protostellar envelopes \citep{Jorgensen2004b,Anderl2016}.

For young protostars, the emission of complex organic molecules (COMs) has been used as tracer of the inner region warm enough to sublimate water ice, the so called hot core, because these species are expected to be trapped in water ice. However, the exact binding energy, and thus spatial extent, will differ for different molecules. Compact COM emission has been detected toward NGC1333-IRAS2A (hereafter IRAS2A) extending between $\sim0\farcs4$ and $\sim1\farcs0$ \citep[e.g.,][]{Jorgensen2005,Maret2014,Maury2014}, similar to the size ($\sim0\farcs8$) of the H$_2^{18}$O emission \citep{Persson2012}. The exact relationship between COMs and the water snowline remains unclear though. For example, the extent of methanol emission in IRAS2A is half the size of the H$_2^{18}$O emission \citep{Maret2014}, while it is twice the size in NGC1333-IRAS4A \citep{Anderl2016}. In addition, \citet{Persson2012} conclude that C$_2$H$_5$CN is likely not related to sublimation of water ice. Since the detection of COMs is very difficult in mature protoplanetary disks, with only methyl cyanide and methanol being observed so far \citep{Oberg2015b,Walsh2016}, they are not suited as tracer of the water snowline in these systems. 

The best candidate to chemically trace the water snowline is HCO$^+$, because its most abundant destroyer in warm dense gas is gaseous H$_2$O:

\begin{equation}
\mathrm{H}_2\mathrm{O} + \mathrm{HCO}^+ \rightarrow \mathrm{CO} + \mathrm{H}_3\mathrm{O}^+.
\end{equation}

\noindent A strong decline in HCO$^+$ is thus expected when H$_2$O desorbs off the dust grains \citep{Phillips1992,Bergin1998}. ALMA observations of the optically thin isotopologue H$^{13}$CO$^+$ toward the Class 0 protostar IRAS 15398–3359 indeed revealed ring-shaped emission \citep{Jorgensen2013}. The spatial distribution is consistent with destruction by water in the innermost region, but the inner radius  of the H$^{13}$CO$^+$ emission is further out than expected. This can be explained if the temperature has recently been higher, that is, if the source has undergone a luminosity outburst. Follow-up observations by \citet{Bjerkeli2016} did not detect the H$_2^{18}$O $4_{14}-3_{21}$ high excitation transition ($E_{up}=322$ K). However, an HDO transition with lower upper level energy ($E_{up}=22$ K) was clearly detected. Although HDO emission is also present in the outflow lobes, the observations are consistent with the H$_2$O-HCO$^+$ anticorrelation scenario.  

Protostellar envelopes are good targets to test the H$^{13}$CO$^+$-H$_2^{18}$O anticorrelation. Because of the higher luminosity (due to higher accretion rates) and the lower vapor pressure, the snowline is located further away from the star than in protoplanetary disks \citep[10s--100 AU instead of a few AU,][]{Harsono2015,Cieza2016}. In addition, H$_2^{18}$O has already been observed toward four of these objects \citep{Jorgensen2010,Persson2012,Persson2013}. The only thing lacking are thus high-spatial resolution images of H$^{13}$CO$^+$. Here, we present NOEMA observations of H$^{13}$CO$^+$ toward one source, IRAS2A, and compare these to the H$_2^{18}$O emission presented by \citet{Persson2012}. IRAS2A is a deeply embedded Class 0 protostar in the NGC1333 region of the Perseus molecular cloud. The quadruple outflows suggest that it is a close binary \citep{Jorgensen2004}, which has recently been confirmed \citep[$0\farcs6$ separation,][]{Tobin2015}. Similar to \citet{Persson2012} we adopt a distance of 250 pc \citep{Enoch2006}.

The observations and results are presented in Sects.~\ref{sec:Observations}~and~\ref{sec:Results}, and compared with the H$_2^{18}$O observations from \citet{Persson2012}. In Sect.~\ref{sec:Discussion}, the integrated H$^{13}$CO$^+$ emission is analyzed using 1D radiative transfer modeling of a spherical envelope with a parametrized H$^{13}$CO$^+$ abundance profile. In addition, this abundance profile is validated against the outcome of a full chemical network and low-resolution DCO$^+$ observations. The main conclusions, including that we confirm the predicted anticorrelation between H$^{13}$CO$^+$ and H$_2^{18}$O, are summarized in Sect.~\ref{sec:Summary}.

\begin{figure*}
\centering
\includegraphics[width=\textwidth,trim={0cm 0.4cm 0cm 0.4cm},clip]{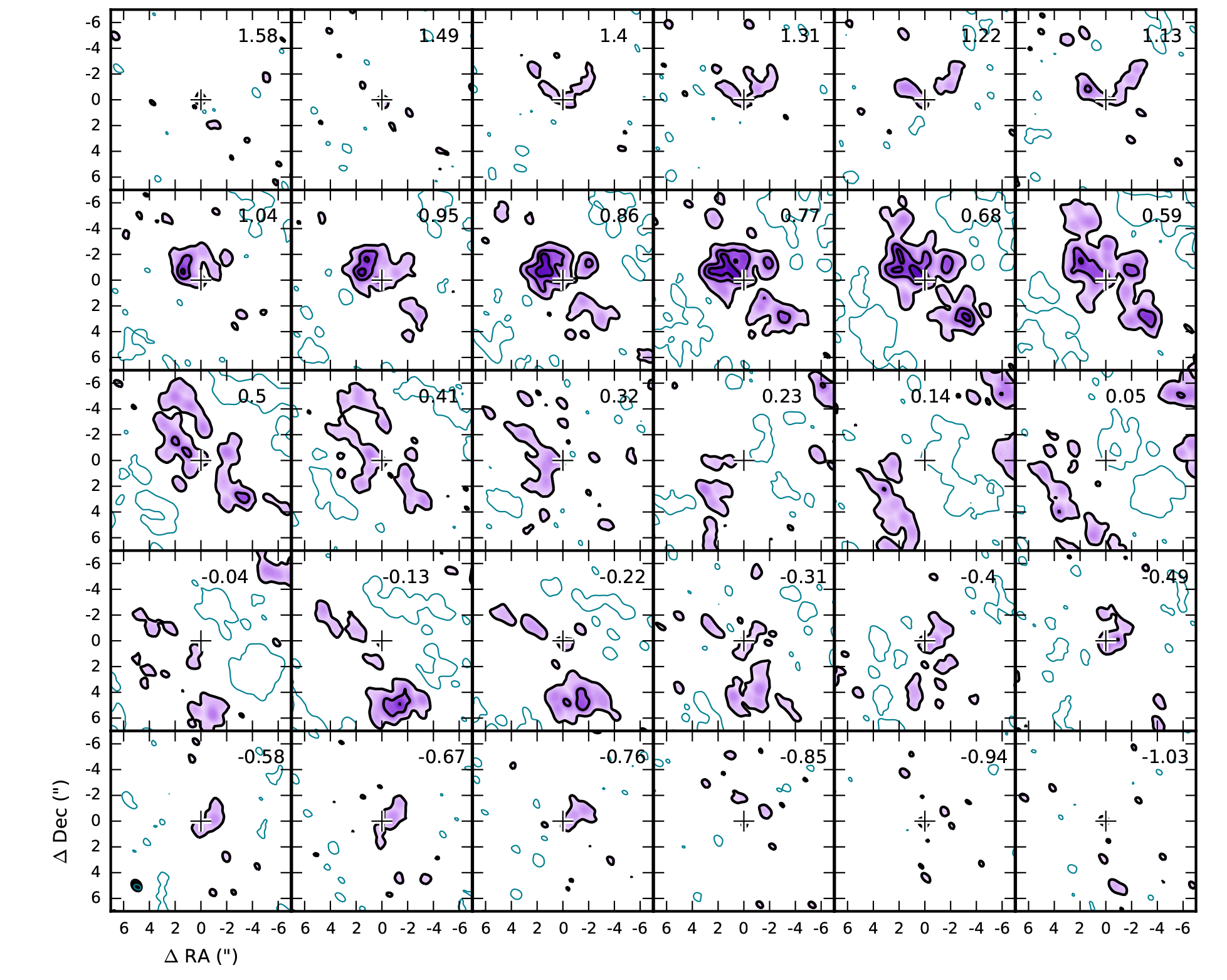}
\caption{Channel maps for the H$^{13}$CO$^+$ $J=3-2$ transition. Black contours are in steps of 10$\sigma$, starting from a 3$\sigma$ level (30~mJy~beam$^{-1}$ in 0.09~km~s$^{-1}$ channels), and the -3$\sigma$ contours are shown in blue. The peak intensity is 324~mJy~beam$^{-1}$ at 7.68~km~s$^{-1}$ ($v_{\rm{lsr}} +$ 0.68~km~s$^{-1}$). The continuum peak position is marked with a cross and the beam is shown in the lower left corner of the bottom left panel. Channel velocities with respect to the systemic velocity of $v_{\rm{lsr}} \approx$~7.0~km~s$^{-1}$ are listed in the top right corner of each panel. For the adopted distance to IRAS2A, $1\arcsec$ corresponds to 250~AU.}
\label{fig:Channelmaps}
\end{figure*}

\begin{figure}
\centering
\includegraphics[width=0.361\textwidth]{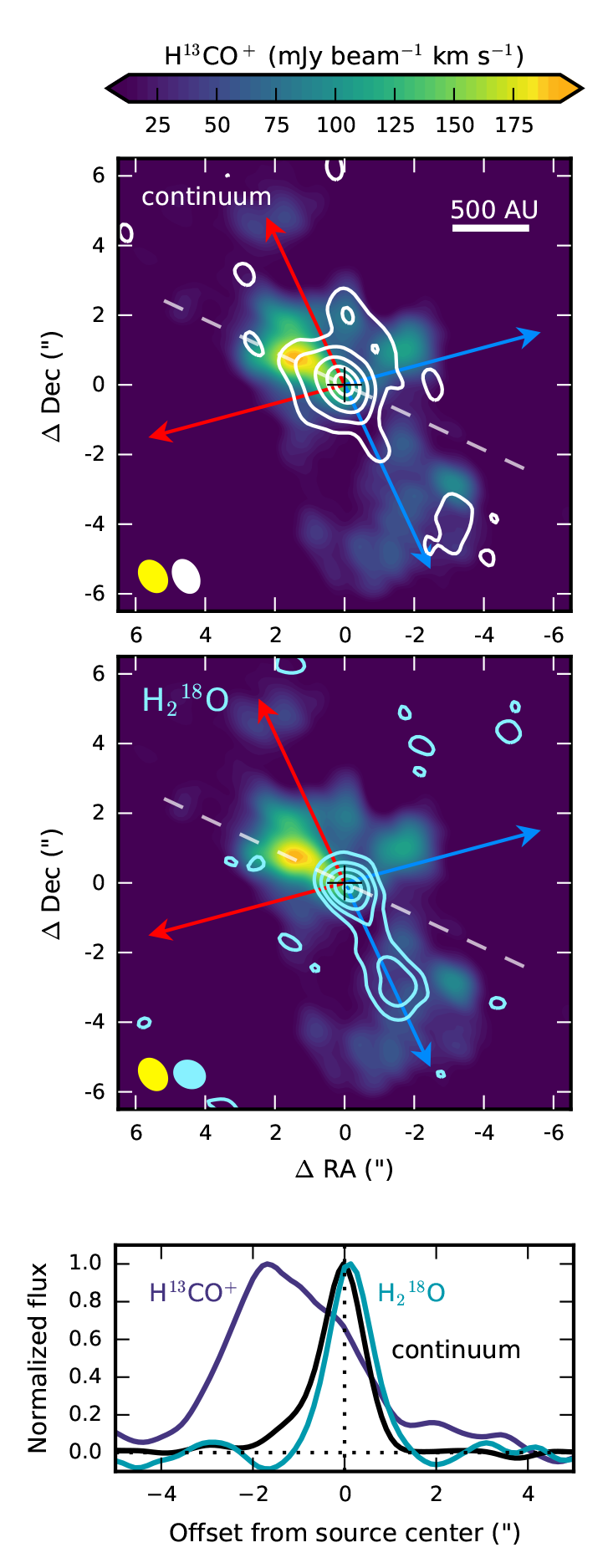}
\caption{Integrated intensity map for the H$^{13}$CO$^+$ $J=3-2$ transition (color scale) toward IRAS2A, with the 1.2 mm continuum overlaid in white contours (\textit{top panel}) and the H$_2^{18}$O $3_{1,3}-2_{2,0}$ transition in blue contours (\textit{middle panel}). The continuum contours are 1.8 (1$\sigma$) $\times$ [3, 10, 25, 50, 75] mJy~beam$^{-1}$, and the H$_2^{18}$O contours are 9.8 (1$\sigma$) $\times$ [3, 8, 15, 25, 35] mJy~beam$^{-1}$~km~s$^{-1}$. The beams are depicted in the lower left corners. The position of the continuum peak is marked by a black cross (the close binary is unresolved in these data) and the outflow axes by red and blue arrows. The integrated intensities along the dashed white line are shown in the \textit{bottom panel}, normalized to their maximum value. The zero flux level and source position are marked by dotted lines. }
\label{fig:Overlay}
\end{figure}

\begin{figure}
\centering
\includegraphics[trim={0.3cm 17.6cm 0cm 0.7cm},clip]{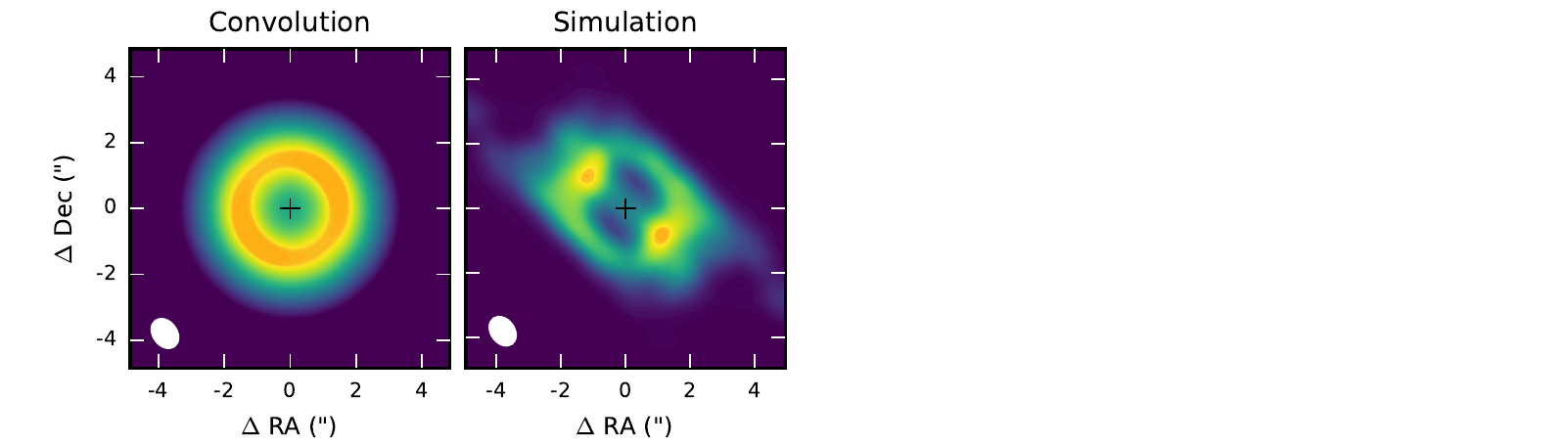}
\caption{Integrated intensity maps for a spherically symmetric ring-shaped H$^{13}$CO$^+$ distribution. Emission simulated with the radiative transfer code Ratran is convolved with the observed beam (\textit{left panel}), and simulated with the ($u,v$) coverage of the NOEMA observations (\textit{right panel}). The southern peak in the simulations is probably not observed due to presence of water in the outflow (see text for details). The continuum peak is denoted with a cross, and the beam is shown in the lower left corner.}
\label{fig:Conv-vs-Sim}
\end{figure}

\begin{figure}
\centering
\includegraphics[trim={0cm 13.5cm 0cm 0.5cm},clip]{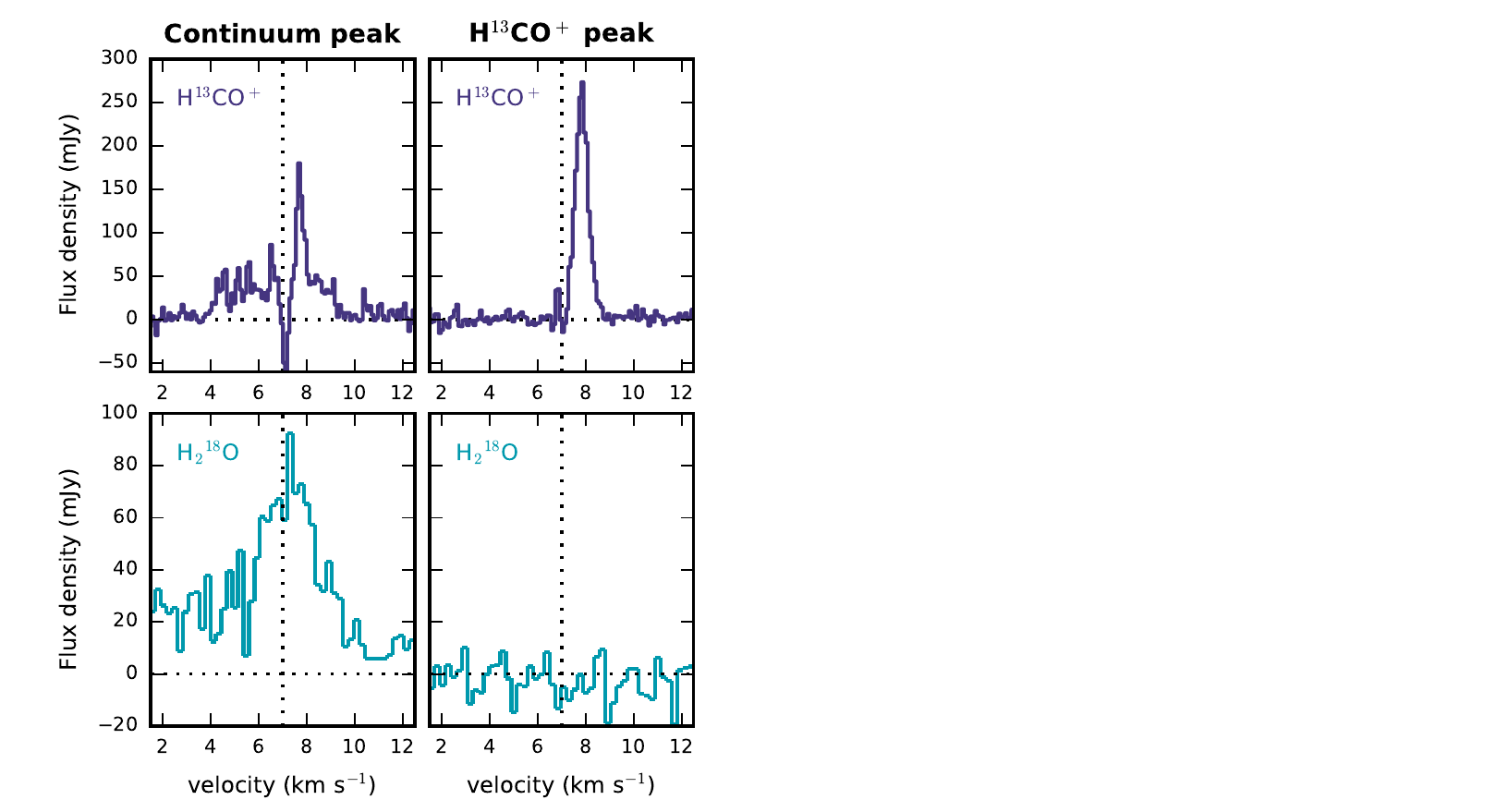}
\caption{Spectra extracted in a beam toward the continuum peak position (\textit{left panels}) and the H$^{13}$CO$^+$ peak position ($\sim2^{\prime\prime}$ NE of the continuum peak, \textit{right panels}). The \textit{top panels} show the H$^{13}$CO$^+$ $J=3-2$ transition and the \textit{bottom panels} the H$_2^{18}$O $3_{1,3}-2_{2,0}$ transition. The two spectra for each species are shown on the same vertical scale. The systemic velocity is marked by the vertical dotted line at $v_{\rm{lsr}}$ $\approx$ 7 km s$^{-1}$.}
\label{fig:Lineprofiles}
\end{figure}


\section{Observations} \label{sec:Observations}

IRAS2A ($\alpha$(2000) = 03$^{\rm{h}}$28$^{\rm{m}}$55$\fs$58; $\delta$(2000) = 31$\degr$14$\arcmin$37$\farcs$10) was observed using the NOrthern Extended Millimeter Array (NOEMA) on December 1 2015 (C configuration), and April 9 2016 (B configuration), for a total of 2.3 h on source in B configuration and 1.2 h in C configuration.  Combining the observations in the different configurations, the data cover baselines from 16.8 to 456.9 m (14.5 to 396.6 k$\lambda$). The receivers were tuned to the H$^{13}$CO$^+$ $J=3-2$ transition at 260.255~GHz (1.15~mm), and the narrow band correlator was set up with one unit (bandwidth of 40 MHz) centered on this frequency providing a spectral resolution of 0.078 MHz (0.09 km~s$^{-1}$). In addition, the WideX correlator was used, covering a 3.6 GHz window (259.2--262.8 GHz) at a resolution of 1.95 MHz (2.2--2.3 km s$^{-1}$). 

\begin{table}
\addtolength{\tabcolsep}{-2pt}
\caption{Overview of the molecular line observations toward IRAS2A.
\label{tab:Lineparameters}} 
\centering
\begin{tabular}{l c c c c}
    \hline\hline
    \\[-.3cm]
    
    Transition & Frequency & $E_{up}/k$ & Beam & $\Delta v$\tablefootmark{a}  \\ 
    & (GHz) & (K) & ($\arcsec$) & (km s$^{-1}$) \\
    \hline 
    \\[-.3cm]
    H$^{13}$CO$^+$ $J=3-2$        & 260.255   & 25  & 0.93$\times$0.68 & 0.09  \\
    H$_2^{18}$O $3_{1,3}-2_{2,0}$ & 203.408   & 204 & 0.87$\times$0.72 & 0.115 \\
    DCO$^+$ $J=2-1$               & 144.068   & 10  & 2.1$\times$1.7   & 4.06  \\
    \hline
\end{tabular}
\tablefoot{\tablefoottext{a}{Velocity resolution.}}
\addtolength{\tabcolsep}{+2pt}
\end{table}

Calibration and imaging were performed using the \texttt{CLIC} and \texttt{MAPPING} packages of the IRAM \texttt{GILDAS} software\footnote{\texttt{http://www.iram.fr/IRAMFR/GILDAS}}. The standard calibration procedure was followed using the quasars 3C454.3 and 3C84 to calibrate the bandpass, 0333+321 to calibrate the complex gains and the point sources MWC349 and LkH$\alpha$101 to calibrate the absolute flux scale. The continuum visibilities were constructed from line-free channels in the WideX spectrum, and the continuum was subtracted from the line data before imaging. Both the line and continuum data were imaged using default robust weighting. For H$^{13}$CO$^+$ this resulted in a $0\farcs93\times0\farcs68$ (PA = 36$^{\circ}$) beam, comparable to the beam of the H$_2^{18}$O observations (see below), and a rms of 10~mJy~beam$^{-1}$ in 0.09 km~s$^{-1}$ channels. The continuum image has a $0\farcs95\times0\farcs65$ (PA = 27$^{\circ}$) beam. The continuum flux obtained from a Gaussian fit to the visibilities is 347~mJy. This is consistent with previous observations between 0.8 and 1.47 mm \citep{Jorgensen2007,Persson2012,Taquet2015} within the 20\% flux calibration error, assuming a power-law spectrum for thermal dust emission ($F_{\nu} \propto \nu^{\alpha}$, with $\alpha \sim 2.0-2.4$). The $(u,v)$-fitted continuum peak position ($\alpha$(2000) = 03$^{\rm{h}}$28$^{\rm{m}}$55$\fs$57; $\delta$(2000) = 31$\degr$14$\arcmin$36$\farcs$92) is located $\sim 0\farcs1$ south of the bright component of the $0\farcs6$ binary identified by \citep{Tobin2015}. 

The H$_2^{18}$O $3_{1,3}-2_{2,0}$ transition at 203.408~GHz (1.47 mm) was observed using the Plateau de Bure Interferometer (PdBI) in December 2009 and March 2010 and was presented by \citet{Persson2012}. The data have a spectral resolution of 0.087 MHz (0.115 km s$^{-1}$) and a resulting beam size of $0\farcs87\times0\farcs72$ (PA = 63.5$^{\circ}$) using natural weighting. 

In addition, IRAS2A was observed using the PdBI in July, August and November 2010 and March 2011 in the C and D configurations as part of a study of complex organic molecules toward low-mass protostars \citep{Taquet2015}. The WideX backends were used at $\sim$145~GHz providing a bandwidth of 3.6~GHz with a spectral resolution of 1.95~MHz \mbox{($\sim 3.5 - 4$~km~s$^{-1}$).} One of the targeted lines was the DCO$^+$ $J=2-1$ transition  at 144.068 GHz. Phase and amplitude were calibrated by performing regular observations of the nearby point sources 3C454.3, 3C84, and 0333+321. Imaging with natural weighting resulted in a $2\farcs1 \times 1\farcs7$ (PA = -155$^{\circ}$) beam. 

An overview of the observed molecular lines and their parameters is provided in Table~\ref{tab:Lineparameters}.


\section{Results} \label{sec:Results}

The H$^{13}$CO$^+$ channel maps presented in Fig.~\ref{fig:Channelmaps} show detection of H$^{13}$CO$^+$ at velocities between 6.2 and 8.4 km~s$^{-1}$ (\mbox{$v_{\rm{lsr}}$ = 7.0 km~s$^{-1}$}, \citealt{Persson2012}), although the emission is brightest for the red-shifted velocities. In none of the channels does the H$^{13}$CO$^+$ emission peak at the source position. Instead, emission peaks are observed at offsets (RA,Dec) of approximately \mbox{(-2$\arcsec$,1.5$\arcsec$)} and \mbox{(3$\arcsec$,-3$\arcsec$)} in the red-shifted channels, and at \mbox{(1$\arcsec$,-6$\arcsec$)} in the blue-shifted channels. H$_2^{18}$O, on the other hand, shows compact emission on source (located by \citealt{Persson2012} $\sim$0$\farcs1$ southwest of the continuum peak) and extends along the southern outflow axis. Overlaying the integrated intensity maps for H$^{13}$CO$^+$ and H$_2^{18}$O (see Fig.~\ref{fig:Overlay}) shows that the H$^{13}$CO$^+$ emission surrounds the water emission and peaks $\sim$2$\arcsec$ northeast of the continuum peak. 

The asymmetry in the spatial distribution of the H$^{13}$CO$^+$ emission is partly due to the sampling of the ($u,v$) plane; simulating emission from a spherical envelope model with a ring-like H$^{13}$CO$^+$ abundance distribution using the same ($u,v$) coverage as the observations results in emission peaks in the northeast and southwest, rather than a spherically symmetric emission pattern (see Fig.~\ref{fig:Conv-vs-Sim}). The absence of a strong southwestern peak in the data may be attributed to the presence of water in the southern outflow. Alternatively, it could be an effect of the three-dimensional structure of the source, which causes the H$^{13}$CO$^+$ emission from that part to be blocked. Finally, it could perhaps be the result of asymmetries in the initial filamentary structure at large scales ($\gg$ 10,000 AU). 

Spectra for H$^{13}$CO$^+$ and H$_2^{18}$O extracted toward the continuum and H$^{13}$CO$^+$ peak positions are presented in Fig.~\ref{fig:Lineprofiles}. Although there is still H$^{13}$CO$^+$ emission present on source, it is reduced by $\sim$40\% compared to the region where water emission is absent. The narrow width (FWHM $\approx$ 0.8 km~ s$^{-1}$) of the H$^{13}$CO$^+$ line at its peak position indicates that the emission does not originate in the outflow. In comparison, the H$_2^{18}$O line width is $\sim$3.5~km~s$^{-1}$. In addition, it has been shown in \citet{Persson2012} that the outflow component of the H$_2^{18}$O emission can be distinguished spatially from the compact component. The emission around $v_{lsr} =$ 3~km~s$^{-1}$ in the H$_2^{18}$O spectrum is due to dimethyl ether (CH$_3$OCH$_3$, \citealt{Persson2012}). 

The absorption feature for H$^{13}$CO$^+$ around the systemic velocity is partly due to emission from the large-scale envelope being resolved out, as can be seen from the channel maps in Fig.~\ref{fig:Channelmaps} where most of the emission is absent around the systemic velocity. In addition, a comparison with JCMT single dish observations from \citet{Jorgensen2004a} shows that only $\sim$7\% of the peak flux is recovered in the NOEMA observations, demonstrating that the bulk of the H$^{13}$CO$^+$ is located on larger scales. Imaging the data before continuum subtraction shows that some absorption also occurs against the continuum in the central channels, indicative of cold gas along the line of sight at scales larger than the scale of interest. 

\begin{figure}
\centering
\includegraphics[trim={.2cm 4.5cm 0cm 1.5cm},clip]{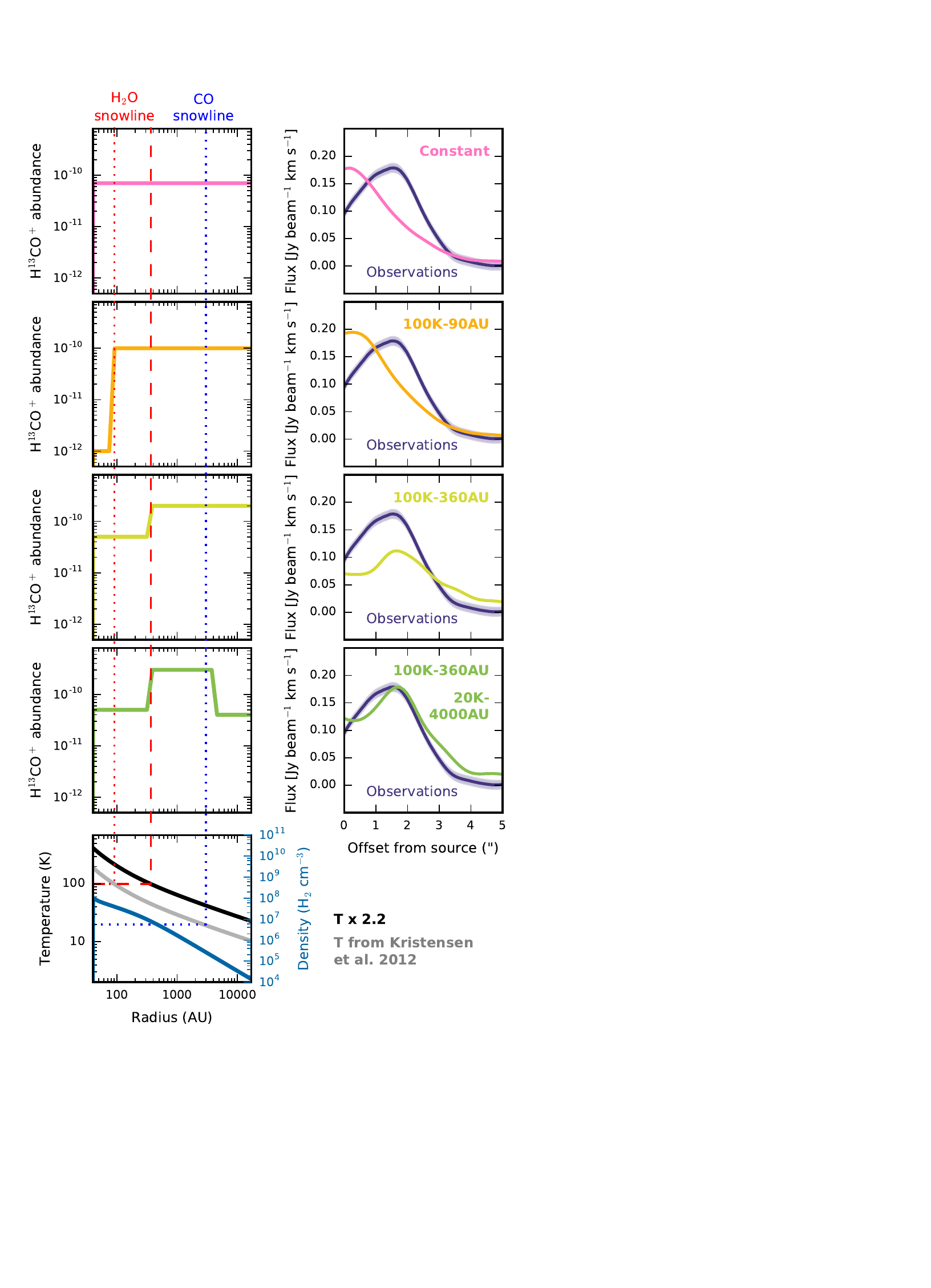}
\caption{Different H$^{13}$CO$^+$ abundance profiles (\textit{left panels}) and the resulting simulated integrated emission along the northeastern part of the radial cut shown in Fig.~\ref{fig:Overlay} (\textit{right panels}). The observed emission is shown in purple. The location of the abundance jumps (in Kelvin and AU) is indicated in the top right corners. \textit{Bottom panel:} Temperature (gray) and density (blue) profiles for the IRAS2A envelope from \citet{Kristensen2012} used in the top two models. The location of the H$_2$O and CO snowlines (at 100 K and 20 K, resp.) are marked by the dotted red and blue lines, respectively. The temperature profile increased by a factor of 2.2, used in the bottom two models, is shown in black. The resulting H$_2$O snowline is indicated by the dashed red line while the CO snowline now falls outside the adopted radial range. For IRAS2A, 1$^{\prime\prime}$ corresponds to $\sim$250 AU.}
\label{fig:RadTransfer}
\end{figure}


\section{Analysis and discussion} \label{sec:Discussion}

\subsection{Parametrized abundance profile for H$^{\sf13}$CO$^+$}

To establish the origin of the dip in the H$^{13}$CO$^+$ emission at the central position, 1D spherically symmetric physical-chemical modeling has been performed using the radiative transfer code Ratran \citep{Hogerheijde2000}. For the physical structure we adopt the temperature and density profiles for IRAS2A from \citet{Kristensen2012}, derived using the 1D spherically symmetric dust radiative transfer code DUSTY \citep{Ivezic1997}. In this procedure the free model parameters were fitted to the spectral energy distribution (SED) and the spatial extent of the sub-mm continuum emission. Based on the chemical consideration that H$^{13}$CO$^+$ is particularly abundant when its main destructor H$_2$O is frozen out, the simplest H$^{13}$CO$^+$ distribution is an abundance profile with a low abundance inside the 100~K radius (i.e., the water snowline) and a higher abundance at larger radii. Both the inner and outer abundances are varied to get the best match to the observed integrated intensity along the radial cut shown in Fig~\ref{fig:Overlay}. The absence of the southwest H$^{13}$CO$^+$ peak (discussed in Sect.~\ref{sec:Results}), cannot be modeled with an axisymmetric 1D model. We therefore restrict the comparison to the northeastern part of the radial cut. For a first analysis, we do not adopt a velocity structure. The difference between the blue- and red-shifted channels can then not be reproduced exactly. Since there is almost no emission along the radial cut in the blue-shifted channels, only the red-shifted channels are used to construct the zeroth moment maps. Including all channels does not significantly change the observed integrated radial profile, but only affects the H$^{13}$CO$^+$ abundance required to reach the observed intensity (by a factor of $\sim$3). However, deriving the precise H$^{13}$CO$^+$ abundance is not the goal of this work. Finally, simulated images representing the observed $(u,v)$ coverage are produced from the Ratran images in \texttt{GILDAS} by creating a $(u,v)$ table with the function \texttt{uv\textunderscore fmodel}. 

\begin{figure*}
\centering
\includegraphics[trim={0cm 12.6cm 0cm 0cm},clip]{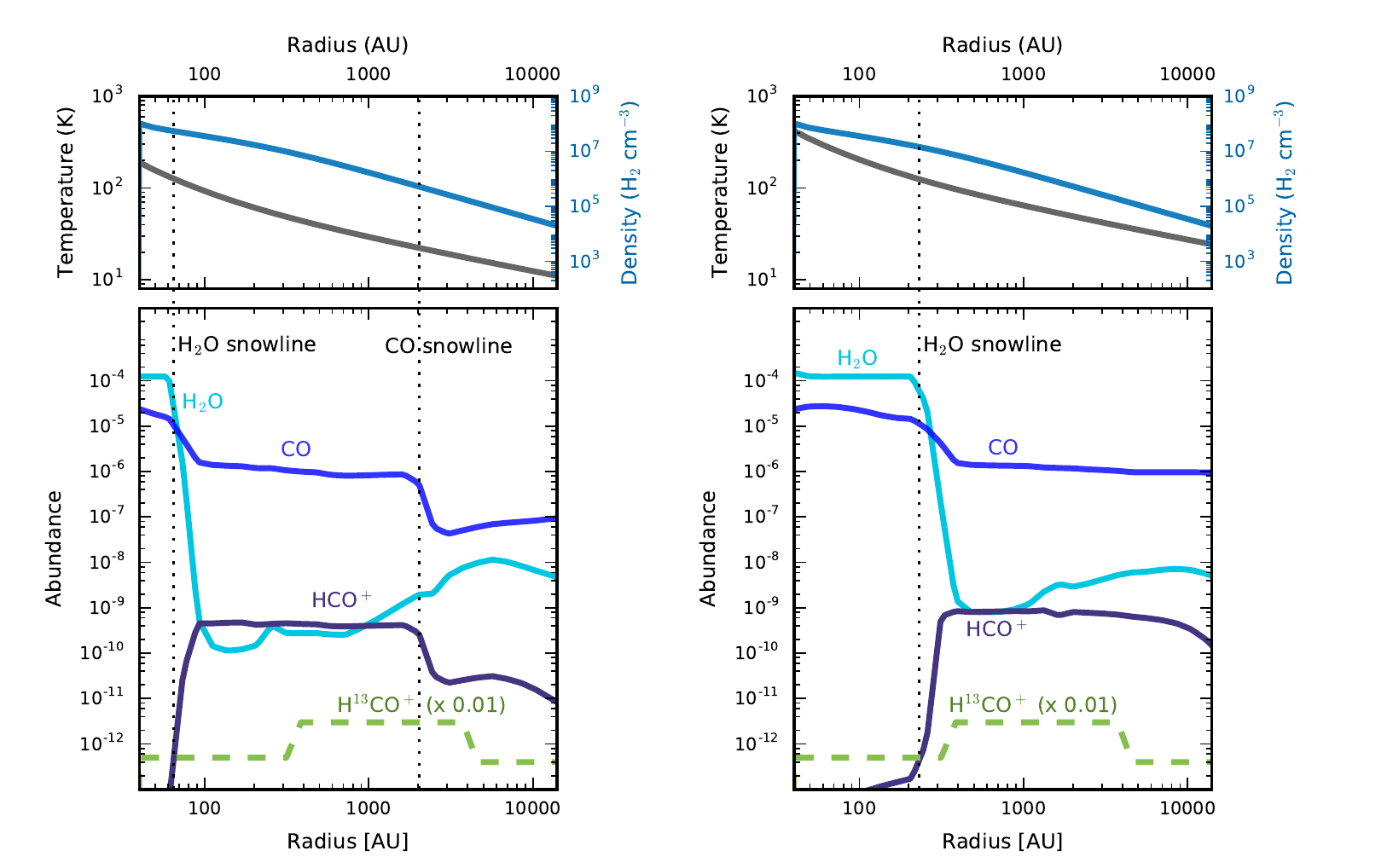}
\caption{\textit{Left panels:} Temperature (gray) and density (blue) profiles for the IRAS2A envelope from \citet{Kristensen2012} (\textit{top}), and the corresponding H$_2$O (light blue), CO (blue) and HCO$^+$ (purple) abundances predicted by the GRAINOBLE model (\textit{bottom}). The dashed green line shows the empirically inferred abundance profile for H$^{13}$CO$^+$ (scaled down by a factor 100). \textit{Right panels:} As left panels, but with the temperature increased by a factor 2.2. The vertical dotted lines mark the H$_2$O and CO snowlines.}
\label{fig:Chemistry}
\end{figure*}

Different H$^{13}$CO$^+$ abundance profiles with their corresponding integrated emission profiles are presented in Fig.~\ref{fig:RadTransfer}. For the adopted physical structure, the temperature drops below 100 K around 90 AU. An H$^{13}$CO$^+$ abundance of $1\times10^{-10}$ at radii $>$90 AU can roughly reproduce the observed integrated intensity (Fig.~\ref{fig:RadTransfer}, second row). However, independent of the inner abundance, the emission always peaks on source, as in the case of a constant H$^{13}$CO$^+$ abundance at all radii (Fig.~\ref{fig:RadTransfer}, top row). Although a large fraction ($\sim$90\%) of the emission is already being resolved out with the ($u,v$) coverage of the observations, this thus does not result in a dip in the H$^{13}$CO$^+$ emission toward the source position. 

To reproduce the location of the emission peak, the abundance jump has to be shifted outward to a radius of $\sim$360~AU. To move the 100~K radius, representing the water snowline, this far out, the temperature profile has to be multiplied by a factor of 2.2 (Fig.~\ref{fig:RadTransfer}, third row). However, the observed integrated intensity cannot be reached with this abundance profile, as the H$^{13}$CO$^+$ emission becomes optically thick for outer abundances higher than $3\times10^{-10}$. Varying the abundance does not affect the position of the emission peak. 

The best match to the observed H$^{13}$CO$^+$ radial profile can be reached by introducing an additional drop in the abundance around $\sim$4000~AU (Fig.~\ref{fig:RadTransfer}, fourth row). An increase in abundance of at least a factor of six is required around 360~AU to get the emission peak at the observed location, but the exact radius of the outer drop does not influence the peak position. Such a decrease in the H$^{13}$CO$^+$ abundance can be expected around the CO snowline because formation of HCO$^+$ will be limited in this region due to a reduction in the parent gaseous CO molecules. A radius of 4000~AU corresponds to 17~K for the \citet{Kristensen2012} temperature profile, consistent with the CO freeze-out temperature for a pure CO ice binding energy. In contrast, for the increased temperature profile the temperature is 38~K at this radius; temperatures below 20~K are not reached within the 20,000 AU radial extent of the model. An increased temperature is thus required to set the inner radius of the H$^{13}$CO$^+$ abundance at the water snowline, while the original temperature places the outer radius around the CO snowline. This discrepancy will be further discussed in Sects.~\ref{sec:Chemmodel} and \ref{sec:DCO+}.  The main result of these models is that the observed position of the H$^{13}$CO$^+$ emission peak can only be reproduced by a reduction in the  abundance of at least a factor 6 in the inner $\sim$360~AU. 

\subsection{Comparison with a full chemical model} \label{sec:Chemmodel}

To inspect the legitimacy of the simple abundance profile, we compare it to the outcome of the astrochemical code GRAINOBLE \citep{Taquet2012b,Taquet2012a,Taquet2013a,Taquet2014}. In short, GRAINOBLE uses the rate equations approach \citep{Hasegawa1992} considering three types of reactions (a so-called three-phase model based on \citealt{Hasegawa1993}): reactions for gas-phase species, reactions for species on the ice surface, and reactions for bulk ice species. The gas-phase chemical network is taken from the KIDA dabatase \citep{Wakelam2015} supplemented by ion-neutral reactions described in \citet{Taquet2016}. A cosmic ray ionization rate of $5\times10^{-17}$~s$^{-1}$ is used. In addition to thermal desorption, UV-photodesorption and chemical desorption upon formation of an exothermic surface reaction have been included as desorption processes. The surface chemical network follows the results of recent laboratory experiments in order to form the main ice species \citep[see][for references]{Taquet2013a}. For CO and H$_2$O binding energies have been adopted for an amorphous water ice substrate: 1150~K and 5775~K, respectively \citep{Collings2004,Fraser2001}. The formation of interstellar ices is first followed during the formation of a dense starless core with the methodology described in \citet{Taquet2014}. This ice chemical composition is then used as initial conditions for the gas-grain chemical evolution around IRAS2A, adopting the static 1D physical profile derived by Kristensen et al. (2012). Surface reactions have a negligible impact on the HCO$^+$ abundance during the latter protostellar phase. They have a significant importance during the former prestellar phase since they govern the transformation of CO ice into other species, thereby setting the initial ice abundance for the protostellar phase. The initial gas-phase abundances and the ice abundances at the start of the protostellar phase are listed in Appendix~\ref{ap:chem}.

The left panel of Fig.~\ref{fig:Chemistry} presents the H$_2$O, CO and HCO$^+$ abundances for the IRAS2A envelope structure after $10^4$ yr of chemical evolution. The abundance profiles do not change significantly for evolution up to $\sim$10$^5$ yr, that is, the duration of the protostellar phase. Going from the outer envelope inward, the temperature rises above 20 K at a radius of $\sim$2700~AU, causing CO to desorb from the ice surface into the gas phase. In the three-phase model, a significant part of the icy CO remains trapped in the ice matrix at temperatures higher than the CO desorption temperature. These CO molecules evaporate together with water when the temperature exceeds $\sim$100~K, resulting in a double abundance jump profile for CO. The water snowline, that is, the radius where 50\% of the water is in the gas phase and 50\% is frozen out, is located at a distance of $\sim$65~AU from the protostar (corresponding to 127 K). As expected, the HCO$^+$ abundance increases sharply at radii beyond the water snowline where its main destructor is frozen out, and decreases again outside the CO snowline where the availability of its parent molecule is reduced. Increasing the temperature by a factor of 2.2, as was required to match the observed H$^{13}$CO$^+$ emission with a drop in the abundance at the 100~K radius, results in a similar chemical profile, only with the snowlines shifted to larger radii (Fig.~\ref{fig:Chemistry}, right panel). The water snowline is now located at $\sim$225~AU (127 K), while the CO snowline falls outside the modeled radial range. This location for the water snowline matches with the observed radial extent of the H$_2^{18}$O emission ($\sim$1$\arcsec$ $\approx$ 250~AU).

For comparison, the best H$^{13}$CO$^+$ abundance profile derived from the data is added to Fig.~\ref{fig:Chemistry} (dashed green line). This inferred H$^{13}$CO$^+$ abundance is higher than expected based on the GRAINOBLE results; similar to the predicted HCO$^+$ abundance instead of a factor $\sim$70 lower. This can most likely be explained by the relatively low CO abundance between the CO and H$_2$O snowlines in the chemical model, which directly results in a low HCO$^+$ abundance. The CO abundance is low because a large fraction of the CO ice is converted into methanol during the prestellar phase (Table~\ref{tab:Init2}). In addition, in the three-phase model, not all CO evaporates inside its snowline, but remains trapped in water ice (Fig.~\ref{fig:Chemistry}). However, it is the overall trend that the HCO$^+$ abundance sharply rises at the H$_2$O snowline and drops at the CO snowline, that is important.

The radius where H$^{13}$CO$^+$ has to increase (360~AU) matches the inner radius of the HCO$^+$ distribution in the chemical network with the increased temperature profile (right panel). The water snowline, however, is located $\sim$135~AU closer to the star, at 225 AU. Similar behavior is seen in models for N$_2$H$^+$; the 50\% reduction in CO, or in this case H$_2$O, at the snowline is not yet enough to diminish the destruction of N$_2$H$^+$ or HCO$^+$, respectively, moving the emission ring outward \citep{Aikawa2015,vantHoff2017}. There are thus two effects at play here: first, the HCO$^+$ abundance is expected to peak outside rather than directly at the water snowline (360 vs. 225 AU). Second, this location for the water snowline is further away from the protostar than expected based on the \citet{Kristensen2012} temperature profile (225 vs. 65 AU).

The outer drop in H$^{13}$CO$^+$ abundance matches less well with the predicted HCO$^+$ distributions. The best agreement is for the original temperature profile (for the adopted CO binding energy of 1150~K), although HCO$^+$ decreases $\sim$2000~AU (2.4~K) further in than the radiative transfer modeling derived from the H$^{13}$CO$^+$ data. Due to the shallow gradient of the temperature profile in the outer envelope, the exact location of the CO snowline is very sensitive to small changes in the thermal structure. The mismatch could therefore be due to uncertainties in the derived temperature. Another important parameter for the snowline location is the assumed binding energy for CO, which is related to the assumed ice structure, that is, whether CO ice is mixed with other molecules or pure. Adopting a lower binding energy, corresponding to a CO ice substrate rather than a H$_2$O ice substrate, would shift the snowline to larger radii. \citet{Anderl2016} investigated the CO binding energy using NOEMA observations of C$^{18}$O and N$_2$H$^+$ toward four young protostars and concluded that the extent of the C$^{18}$O emission was best reproduced using a CO binding energy of 1200~K, corresponding to a water ice substrate. In addition, they showed that the freeze-out temperature is $\sim$6~K lower assuming a binding energy of 855~K for pure CO ice instead of 1200~K. Since the temperature difference between the CO snowline and the location of the H$^{13}$CO$^+$ drop is only 2.4~K, a high CO binding energy seems also best for IRAS2A. The small discrepancy between the CO snowline location predicted by the chemical model and the location inferred from the H$^{13}$CO$^+$ observations may then be the result of asymmetries in IRAS2A, as hinted at by the asymmetric C$^{18}$O emission shown by \citet{Anderl2016} for other protostars. 

As noted before, the exact location of the drop in H$^{13}$CO$^+$ in the outer envelope does not influence the observed position of the emission peak. The main point is that the chemical model indeed predicts a steep rise in HCO$^+$ abundance just outside the water snowline, and a drop outside the CO snowline. This justifies the abundance profile used in the radiative transfer modeling, and corroborates that the depression in H$^{13}$CO$^+$ in the center is due to the presence of gas-phase water. 

\subsection{Tracing the CO snowline: signs of an accretion burst?} \label{sec:DCO+}

Having the snowlines located further out than expected based on the adopted temperature profile derived from SED models, such as found here for the water snowline, could be a sign that the protostar has recently undergone a luminosity outburst. Such an outburst heats up the envelope, causing ices to sublimate off the grains. After the burst, the envelope cools rapidly \citep{Johnstone2013}, while it takes much longer for the molecules to freeze back onto the dust grains \citep{Rodgers2003}. As a result, snowlines are shifted away from the star \citep{Visser2015,Jorgensen2015,Cieza2016,Frimann2017}. An increased temperature structure is required to fit the H$^{13}$CO$^+$ emission, suggesting that the water snowline is further out than expected. In case of an outburst, the CO snowline would also be shifted outward. Observational constraints for the CO snowline can thus help to establish whether IRAS2A has indeed recently undergone an accretion burst.  

\begin{figure}
\centering
\includegraphics[trim={0cm 15.2cm 0cm 0.6cm},clip]{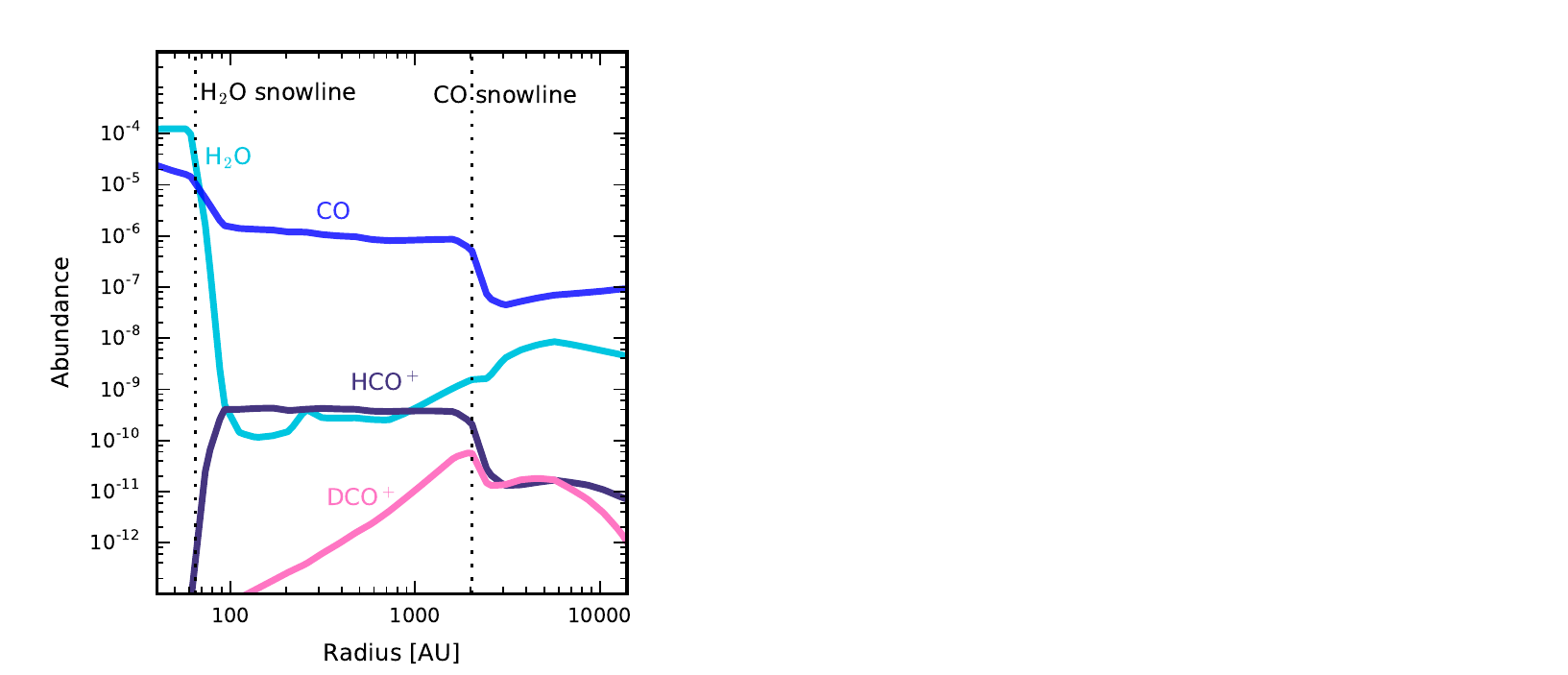}
\caption{H$_2$O (light blue), CO (blue), HCO$^+$ (purple) and DCO$^+$ (pink) abundances predicted by the GRAINOBLE model for the IRAS2A temperature and density profile from \citet{Kristensen2012}. The vertical dotted lines mark the H$_2$O and CO snowlines.}
\label{fig:DChemistry}
\end{figure}

The deuterated form of HCO$^+$, DCO$^+$, can be used to trace the the freeze-out of CO (e.g., \citealt{Mathews2013}). DCO$^+$ forms via reaction of H$_2$D$^+$ and CO. Formation of H$_2$D$^+$ is enhanced at low temperatures ($<$30~K), but below $\sim$20~K CO starts to freeze-out. This balance results in DCO$^+$ peaking around the CO snowline \citep{Murillo2015}. Contamination of the emission by DCO$^+$ formed in warmer layers from CH$_2$D$^+$ is not likely to be important in protostellar envelopes, unlike in protoplanetary disks \citep{Favre2015}. The DCO$^+$ abundance profile for the IRAS2A envelope structure is predicted with GRAINOBLE by extending the chemical network with a small network that includes reactions between CO and H$_2$D$^+$, HD$_2^+$ and D$_3^+$. The DCO$^+$ abundance indeed peaks around the CO snowline (see Fig.~\ref{fig:DChemistry}), which is located at $\sim$2000~AU for the fiducial temperature structure. As such, DCO$^+$ emission can be used to estimate the CO snowline location and constrain the temperature profile in the envelope. 

\begin{figure}
\centering
\includegraphics[width=0.345\textwidth]{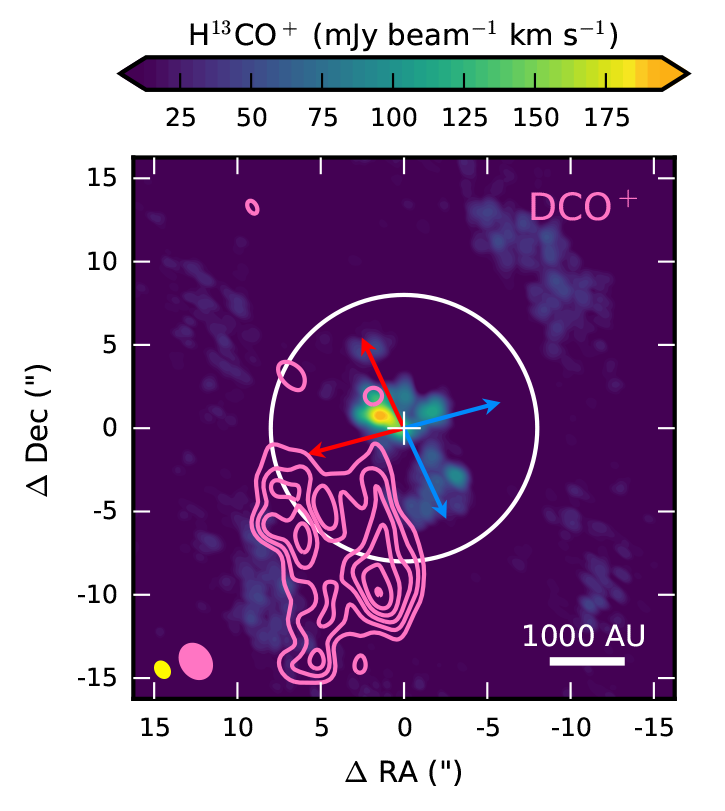}
\caption{Integrated intensity map for the H$^{13}$CO$^+$ $J=3-2$ transition (color scale) toward IRAS2A, with the DCO$^+$ $J=2-1$ transition overlaid in pink contours. Contours are in steps of 1$\sigma$ starting from a 3$\sigma$ level (15~mJy~beam$^{-1}$~km~s$^{-1}$). The beams are depicted in the lower left corner. The position of the continuum peak is marked by a white cross and the outflow axes by red and blue arrows. The length of the arrows is the same as in Fig.~\ref{fig:Overlay} to compare the scales. The solid white contour represents the CO snowline radius predicted by the chemical model for the \citet{Kristensen2012} temperature profile ($\sim$2000~AU).}
\label{fig:DCO+}
\end{figure}

The observed integrated intensity map for DCO$^+$ toward IRAS2A is presented in Fig.~\ref{fig:DCO+}. Emission is observed south of the continuum peak, and peaks $\sim$10$\arcsec$ off source. Such an asymmetric emission pattern is observed for more protostars \citep[e.g.,][]{Murillo2015}. A CO snowline around $\sim$2000 AU, as predicted by the chemical model for the \citet{Kristensen2012} temperature profile, is in agreement with the DCO$^+$ emission. Thus, although the H$_2$O snowline is located at larger radii than predicted based on the IRAS2A temperature profile (225 vs. 65 AU), the location of the CO snowline is as expected. Since the chemical reset after an outburst will be faster at high densities, it is unlikely that the CO snowline has already shifted back while the H$_2$O snowline has not. In addition, \citet{Jorgensen2015} present C$^{18}$O images obtained with the Submillimeter Array (SMA) for which the spatial extent matches well with the current luminosity. It is thus unlikely that IRAS2A has recently undergone an accretion burst. Why the temperature profile matches with the snowline location for CO but not for H$_2$O will be further discussed in Sect.~\ref{sec:physstructure}. 

\subsection{Attenuation of the cosmic ray ionization rate}

The chemistry leading to the formation of HCO$^+$ is initiated by the ionization of H$_2$ molecules by cosmic rays (CRs). A reduction in the CR ionization rate will therefore result in a lower HCO$^+$ abundance. Numerical simulations performed by \citet{Padovani2013} show that the CR ionization rate can be reduced by an order of magnitude in the inner region of a collapsing cloud core due to twisting of the magnetic field lines. However, the exact value of the CR ionization rate in the inner regions of protostellar envelopes is unknown, and more recent work by \citet{Padovani2015, Padovani2016} shows that jet shocks and protostellar surface shocks can drive particle acceleration, leading to increased CR ionization rates. 

Nonetheless, to assess if the effect of CR attenuation is large enough to cause the drop in H$^{13}$CO$^{+}$ emission that we observe, we performed an order of magnitude calculation. As such, we approximate the reactions producing H$_3^+$ upon CR ionization by 
\begin{equation}
  \rm{\zeta_{CR} + H_2} \rightarrow \rm{H_3^+},
\end{equation}
where $\zeta_{\rm{CR}}$ is the cosmic ray ionization rate. In this formalism we ignore H$_2^+$, as the ionization is the rate limiting step for H$_3^+$. Formation of HCO$^+$ is dominated by the reaction between H$_3^+$ and CO,
\begin{equation}
  \rm{H_3^+ + CO} \rightarrow \rm{HCO^+},
\end{equation}
and to estimate the importance of CRs, only recombination is considered as destruction reaction for HCO$^+$: 
\begin{equation} \label{eq:destruction}
  \rm{HCO^+ + e^-} \rightarrow \rm{CO + H}. 
\end{equation}
Assuming steady state and $n$(e$^-$) $\approx n$(HCO$^+$), the following relation can then be derived between the HCO$^+$ density, $n(\rm{HCO^+})$, and the CR ionization rate, $\zeta_{\rm{CR}}$ (see Appendix~\ref{ap:CR} for more details): 
\begin{equation}
  n(\rm{HCO^+}) = \sqrt{\frac{\zeta_{\rm{CR}} n(\rm{H_2})}{k}}, 
\end{equation}
where $n$(H$_2$) is the molecular density of H$_2$ and $k$ is the reaction rate coefficient for the HCO$^+$ destruction reaction (Eq.~\ref{eq:destruction}). From this it can be seen that an order of magnitude reduction in the CR ionization rate gives a factor of three reduction in HCO$^+$ abundance. Based on our radiative transfer modeling, at least a factor of six is required to fit the observations. Attenuation of the CR ionization rate alone is thus not likely to be the cause of the observed reduced H$^{13}$CO$^+$ emission toward the center of IRAS2A.  

\subsection{Physical structure} \label{sec:physstructure}

Finally, a more likely explanation for the discrepancy between the temperatures needed for the CO and H$_2$O snowlines may be that the physical structure of IRAS2A is more complicated than assumed with the 1D model. For several young protostars, it has been shown that a spherical envelope model cannot reproduce all of the continuum emission on small scales \citep[e.g.,][]{Hogerheijde1999,Looney2003}. This is also the case for IRAS2A \citep{Brown2000,Jorgensen2005} and \citet{Persson2016} derive the presence a compact disk-like structure with a 167$^{+63}_{-21}$ AU radius. The density may therefore change significantly on these scales of disk-envelope transition, that is, the scale where we observe the H$_2^{18}$O-H$^{13}$CO$^+$ anticorrelation. Since the envelope around IRAS2A is relatively optically thick to its own infrared radiation on $\sim$100~AU scales, the exact temperature distribution becomes quite dependent on the assumption about the density structure. In addition, the presence of an outflow cavity can shift the 100~K radius outward in specific directions \citep{Bjerkeli2016}, as the temperature along the cavity wall is higher at certain radii than in the envelope midplane \citep{Visser2012}. However, these considerations do not affect the main conclusion that the observations show a clear anticorrelation between H$_2^{18}$O and H$^{13}$CO$^+$ emission.


\section{Summary and outlook}  \label{sec:Summary}

We have presented subarcsecond resolution observations of the H$^{13}$CO$^+$ $J=3-2$ transition at 260.255 GHz toward the Class~0 protostar IRAS2A, and for the first time, compared these directly to observations of H$_2^{18}$O, which were presented by \citet{Persson2012}. While the H$_2^{18}$O $3_{1,3}-2_{2,0}$ emission is centered at the continuum peak within a $\sim$1$\arcsec$ radius, the H$^{13}$CO$^+$ emission peaks $\sim$2$^{\prime\prime}$ to the northeast. 

Using a 1D envelope model for IRAS2A, this offset in H$^{13}$CO$^+$ emission can be explained by an abundance decrease of at least a factor of six in the inner $\sim$360~AU, but not by a constant H$^{13}$CO$^+$ abundance. An increase in H$^{13}$CO$^+$ abundance just outside the water snowline is consistent with chemical model predictions; the water snowline is then located around 225~AU, whereas the H$^{13}$CO$^+$ abundance peaks $\sim$135~AU further out at 360 AU. This snowline radius is larger than expected if the 1D spherically symmetric temperature profile from \citet{Kristensen2012} is adopted. DCO$^+$ observations do place the CO snowline at the expected location, making an outburst scenario unlikely. 

The combined observations of H$^{13}$CO$^+$ and H$_2^{18}$O, together with radiative transfer and chemical modeling, thus provide proof of concept for the HCO$^+$-H$_2$O anticorrelation. The next step would be to corroborate this for more protostellar systems. Establishing H$^{13}$CO$^+$ as a good tracer of the water snowline may allow localization of the snowline in protoplanetary disks for which direct detection is very difficult.


\begin{acknowledgements} 
We would like to thank the referee for useful comments. In addition, we are grateful to the IRAM staff, in particular Miguel Montarg{\`e}s, for his help with the observations and reduction of the data, and Nadia Murillo and Arthur Bosman for useful discussions. Astrochemistry in Leiden is supported by the European Union A-ERC grant 291141 CHEMPLAN, by the Netherlands Research School for Astronomy (NOVA) and by a Royal Netherlands Academy of Arts and Sciences (KNAW) professor prize. M.L.R.H acknowledges support from a Huygens fellowship from Leiden University. M.V.P. postdoctoral position is funded by the ERC consolidator grant 614264. V.T. has received funding from the European Union's Horizon 2020 research and innovation programme under the Marie Sklodowska-Curie grant agreement n. 664931. The research of J.K.J. is supported by European Union ERC Consolidator Grant ``S4F'' (grant agreement No~646908) and Centre for Star and Planet Formation funded by the Danish National Research Foundation.
\end{acknowledgements}


\bibliographystyle{aa} 
\bibliography{H13CO+}


\begin{appendix}


\section{Initial abundances in the GRAINOBLE model} \label{ap:chem}

Table~\ref{tab:Init1} lists the initial elemental abundances in the gas phase for the prestellar phase of the GRAINOBLE model, based on the low-metal model from \citet{Wakelam2008}. The ice composition at the end of this phase (after 1.7~Myr) is listed in Table~\ref{tab:Init2}. These are the initial conditions for the chemical evolution around IRAS2A.  

\begin{table}
\addtolength{\tabcolsep}{-2pt}
\caption{Initial gas-phase elemental abundances.
\label{tab:Init1}} 
\centering
\begin{tabular}{l c}
    \hline\hline
    \\[-.3cm]
    Species & Abundance\tablefootmark{a}  \\ 
    \hline 
    \\[-.3cm]
    H$_2$ 	& 0.5    \\
    HD	 	& 1.5$\times$10$^{-5}$ \\
    H 		& 1.6$\times$10$^{-4}$    \\
    He		& 9.0$\times$10$^{-2}$     \\
    C+		& 7.3$\times$10$^{-5}$     \\
    N		& 2.1$\times$10$^{-5}$     \\
    O		& 1.8$\times$10$^{-4}$     \\
    Si		& 8.0$\times$10$^{-9}$     \\
    S		& 8.0$\times$10$^{-8}$     \\
    Fe		& 3.0$\times$10$^{-9}$     \\
    Na		& 2.0$\times$10$^{-9}$     \\
    Mg		& 7.0$\times$10$^{-9}$     \\
    Cl		& 1.0$\times$10$^{-9}$     \\
    \hline
\end{tabular}
\tablefoot{\tablefoottext{a}{Abundances with respect to the total number of hydrogen nuclei.}}
\addtolength{\tabcolsep}{+2pt}
\end{table}

\begin{table}
\addtolength{\tabcolsep}{-2pt}
\caption{Ice chemical composition at the end of the prestellar phase.
\label{tab:Init2}} 
\centering
\begin{tabular}{l c}
    \hline\hline
    \\[-.3cm]
    Species & Abundance\tablefootmark{a}  \\ 
    \hline 
    \\[-.3cm]
    H$_2$O 	& 1.1$\times$10$^{-4}$     \\
    CO		& 1.3$\times$10$^{-5}$    \\
    CO$_2$	& 1.7$\times$10$^{-7}$     \\
    N$_2$	& 9.8$\times$10$^{-6}$     \\
    H$_2$CO	& 6.7$\times$10$^{-5}$     \\
    CH$_3$OH	& 5.5$\times$10$^{-5}$     \\
    NH$_3$	& 1.3$\times$10$^{-6}$     \\
    CH$_4$	& 4.4$\times$10$^{-6}$     \\
    \hline
\end{tabular}
\tablefoot{\tablefoottext{a}{Abundances with respect to the total number of hydrogen nuclei.}}
\addtolength{\tabcolsep}{+2pt}
\end{table}


\section{Effect of an attenuated cosmic ray ionization rate} \label{ap:CR}

A first order approximation of the effect of cosmic ray ionization rate on the HCO$^+$ abundance can be made by considering the main formation and destruction reactions of HCO$^+$. Assuming steady state, that is, assuming that the HCO$^+$ abundance does not change over time, we can write  
\begin{equation} \label{eq:ode}
  \frac{\mathrm{d}n\rm{(HCO^+)}}{\mathrm{d}t} = \rm{formation \hspace{0.1cm} rate - destruction \hspace{0.1cm} rate} = 0. 
\end{equation}
The main formation pathway of HCO$^+$ is through reaction between CO and H$_3^+$, 
\begin{equation}
  \rm{H_3^+ + CO} \rightarrow \rm{HCO^+},
\end{equation}
and recombination with electrons is the main destruction route,
\begin{equation} \label{eq:hco+formation}
  \rm{HCO^+ + e^-} \rightarrow \rm{CO + H}. 
\end{equation}
The reaction of HCO$^+$ with H$_2$O is ignored in this calculation to assess the importance of the CR ionization rate. 

The rate of these reactions can be written as 
\begin{equation}
  \mathrm{rate} = kn(\mathrm{A})n(\mathrm{B}),
\end{equation}
where $k$ is the rate coefficient, and $n$(A) and $n$(B) are the molecular densities of the reacting species. Eq.~\ref{eq:ode} then becomes 
\begin{equation} \label{eq:hco+}
  k_1n(\mathrm{H_3^+})n(\mathrm{CO}) - k_2n(\mathrm{HCO^+})n(\mathrm{e^-}) = 0. 
\end{equation}

The H$_3^+$ density, $n$(H$_3^+$), can be calculated by following the same procedure. The formation reactions are 
\begin{equation} 
  \rm{H_2} + \zeta_{\rm{CR}} \rightarrow \rm{H_2^+ + e^-},\hspace{0.2cm} \mathrm{and}
\end{equation}
\begin{equation}
  \rm{H_2^+ + H_2} \rightarrow \rm{H_3^+ + H}. 
\end{equation}
In order to solve Eq.~\ref{eq:hco+} analytically, we summarize the formation of H$_3^+$ as follows 
\begin{equation}
  \rm{\zeta_{CR} + H_2} \rightarrow \rm{H_3^+},
\end{equation}
because the ionization reaction is the rate limiting step. In addition, only the reaction with CO (Eq.~\ref{eq:hco+formation}) is considered as destruction mechanism. We can then write an equation similar to Eq.~\ref{eq:hco+}: 
\begin{equation} \label{eq:H3+}
  \zeta_{\mathrm{CR}}n\mathrm{(H_2)} - k_1n(\mathrm{H_3^+})n(\mathrm{CO}) = 0.
\end{equation}

Substituting Eq.~\ref{eq:H3+} into Eq.~\ref{eq:hco+}, and assuming \mbox{$n$(e$^-) \approx n$(HCO$^+$)}, we can solve Eq.~\ref{eq:hco+} to get an expression for the HCO$^+$ density: 
\begin{equation}
  n(\rm{HCO^+}) = \sqrt{\frac{\zeta_{\rm{CR}} n(\rm{H_2})}{k_2}}. 
\end{equation}

\end{appendix}

\end{document}